# Non-s wave superconductivity in boron-doped nanodiamond films with 0-π Josephson junction array

Somnath Bhattacharyya [1,2*], Christopher Coleman [1,2], Davie Mtsuko [1], and Dmitri Churochkin [1]

[1]Nano-Scale Transport Physics Laboratory, School of Physics, [2]DST/NRF Centre of Excellence in Strong Materials, University of the Witwatersrand, Private Bag 3, WITS 2050, Johannesburg, South Africa

Superconducting transport properties of granular materials are greatly influenced by the microstructure. We show that in heavily boron-doped diamond films (HBDDF) films some sharp transport features can be manipulated by applying a magnetic field and controlled finite bias current. We demonstrate the conductivity cross-over from dirty metal to the superconducting state through an insulating peak arising at a very low current or magnetic field region and particularly pronounced negative magnetoresistance with periodic oscillatory features. The current-voltage characteristics show features of the Berezinskii-Kosterlitz-Thouless (BKT) phase transitions which verifies the two-dimensional structure in HBDDF observed recently. A zero bias conductance peak can be attributed to the Andreev bound state formed at the grain boundaries of diamond nanocrystals. The set of observations can be qualitatively explained consistently through the concept of a superconducting transition with a non-$s$ wave order parameter in the diamond heterostructures.

## I. INTRODUCTION

Over several years HBDDF have generated interest in fundamental studies as a new class of disordered superconductor, in which the superconducting diamond grains are separated by an ultra-thin layer of $sp^2$-C hybridised grain-boundary regions [1] [2] [3]. Potential disorder at the randomly distributed boron centres within the individual grain may yield insight to an extreme case in the study of weak localization (WL) and flux pinning and has been linked to new quantum phase [4]. The influence of boron doping on HBDDF has been studied across the metal-insulator transition into the superconducting regime, however, more studies are required to improve the understanding of the exact pairing mechanism [1], [5-14].

It is also proposed that the grain boundaries can act as weak links coupling two neighbouring superconducting grains forming a non-uniform array of Josephson junctions [15]. In order to find the nature of the superconducting gap state scanning tunneling spectroscopy has been performed which showed broadening of the gap, believed to be related to disorder as well as to distribution of grains with different transition temperatures ($T_c$) [3][9]. Temperature dependent studies of the gap showed that while the differential conductance could be fitted in some regions with Bardeen-Cooper-Schrieffer (BCS) theory for weak coupling, this theory broke down in other regions [10] [11]. Theoretical studies employing the standard BCS theory anticipate an exponentially increasing $T_c$ with boron concentration while experimental studies show saturation of the $T_c$ with boron concentration [12], pointing to the presence of additional mechanism limiting the $T_c$. The possibility of a new form of high $T_c$ superconductor based on the resonant states in the valence band stimulated the community's interest in superconductivity in boron doped diamond (films) [13]. A pairing mechanism based on spin-flip interactions of weakly localized orbits has been put forward, however it lacks a detailed theoretical analysis [14]. Vortex behaviour in boron-doped single crystal diamond has also been investigated [11] however, the phase change from bound vortex-antivortex pairs through to unbound vortices has not yet been discussed to check for the possibility of the so-called BKT transitions which has been claimed in networks of 1D single walled carbon nanotubes and 2D graphene-metal hybrids [16] [17] [18].

In recent time this mechanism has been studied thoroughly in bilayer graphene where non-$s$ type order parameter has been suggested [19] [20] [21] [22]. Our approach to qualitatively interpret the transport phenomena in HBDDF is based on the superlattice model utilized for the description of the electron transport in the nitrogen doped nanocrystalline diamond (NCD) films [23]. Recently, the validity of heterostructure approach to the modelling of the transport properties in the HBDDF has been supported by both Raman and x-ray structural analysis of the HBDDF [24]. It is claimed that the superconductivity in HBDDF can be recognized only if small bilayers formed by boron doping either inside the dielectric diamond or on its surfaces. In other words, the interface character of the superconductivity in the HBDDF has been claimed. The presumed strongly anisotropic 2D character of the internal structure of the HBDDF is expected to induce some well-established features, such as; the BKT transition, secondly anisotropy of the order parameter ($\Delta$), and as a consequence the type of pairing can be similar to the case of the strongly anisotropic superconductors, i.e. mainly high $T_c$ and finally the granular character of bilayers makes the Josephson junction (JJ) array interpretation valuable with respects to the description of the tunnelling phenomena in the transport of the HBDDF. It introduces the natural scale (grain size) which can define the localization phenomena and decoherence in the condensate of the Cooper pairs leading to the superconductor-insulator transition [25].

Results of experiments presented in the paper support the above-mentioned rich physics expected for the HBDDF treated as a series of 2D heterostructures [23]. Therefore, we are investigating the zero bias conductance peak (ZBCP) along with BKT signatures in these samples. The former usually attributed to the Andreev bound states (ABS) formed at the interfaces of superconductors with non-$s$ type pairing. The latter, indicates the 2D structural features of the system as well as the importance of transitions in the system of vortices that can be formed in the HBDDF below $T_c$.

Moreover, we study the pronounced current or field induced superconductor insulator transition in HBDDF which shows similarity to what has been observed in thin granular Pb films and interpreted as a consequence of superconductor-insulator tunnelling in the highly resistive regime [25]. The Cooper pair is supposed to be localized on the grain size scale. It motivates us to interpret all observed transport features through the concept of inhomogeneous 2D ensemble of Josephson junctions which are affected by the creation of vortex-antivortex pairs due to the influence of current ($I$) or bias ($V$), magnetic field ($B$), and temperature ($T$).

## II. EXPERIMENTAL DETAILS

The HBDDF samples (B1 and B5) were grown using microwave plasma enhanced chemical vapour deposition. The achieved boron concentrations were 2.8 and $2.0 \times 10^{21}$ cm$^{-3}$ (i.e. well above the Mott metallic transition $\sim 3 \times 10^{20}$ cm$^{-3}$) for two samples grown using 99% and 95% of CH$_4$ in H$_2$, respectively with 4000 ppm trimethylborane (TMB) to CH$_4$. The substrate temperature was 850°C and the pressure was $\sim$ 80 Torr. The microwave power applied was 1.4 kW. Both samples were $\sim$ 420-450 nm thick [14]. While the 5% sample (B5) was fine grained with grains of size $\sim$ 20-30 nm, the sample (B1) prepared from 1%CH$_4$ in H$_2$ was composed of larger grains $\sim$ 50-70 nm in size. Contacts were made using a silver paint in a van der Pauw configuration, allowing for the measurement of both longitudinal and transverse transport properties. Electrical transport measurements were performed using a cryogen free measurement system at temperatures ranging from 0.3 K to 5 K and magnetic fields between 0 and 5 T. This involved a lock in amplifier technique using a Keithly nano-voltmeter.

## III. RESULTS AND DISCUSSION

### A. Resistance *vs.* temperature

We first consider the sample (B5) prepared with 5%CH$_4$ in the chamber (having a smaller grain size than B1) whose resistance ($R$) decreases rapidly with $T$ from around 2 K which can be tuned by the applied field at a high current ($I$ > 100 µA) ((Fig. 1(a)). The critical field is $\sim$ 2.5 T for B5 recorded at $I$ = 10 µA. Above the onset temperature the change in resistance with temperature is very small, however as already observed it decreases towards higher temperatures, around 10 Ω over a 100 K temperature range. The resistance was measured in van der Pauw configuration. The longitudinal resistance ($R_{XX}$) was measured by applying the current across one edge and measuring voltage drop across the opposite edge. We also measured the transverse resistance ($R_{XY}$) by applying a bias current across two diagonally opposite edges and voltage across the remaining two diagonally opposite edges. $R(T)$s were also taken at zero magnetic field for sample B1 at low current $I$ = 28 µA in both $R_{XX}$ and $R_{XY}$ configurations (Fig. 1(b) and inset). Anomalous humps and steps can be observed from *R-T* of sample B1. Lower bias currents resulted in $R(T)$s where the superconducting phase shows negative transverse and longitudinal resistance. It should be noted that similar steps have been observed by other groups and attributed to either granular character of the HBDDF or to the appearance of the new phase because of the boron doping [4]. The samples also exhibited apparent negative resistance under certain measurement conditions, this phenomena was particularly pronounced in the transverse resistance as shown in Fig. 1 (a), inset. The manifestation of negative resistance has before been observed in superconducting systems and was initially explained in terms of vortex backflow due to thermally excited quasiparticles as well as guided vortex motion due to the layered structure of the high T$_c$ materials that first demonstrated this effect [26]. However as the phenomena was later observed in a range of materials including conventional isotropic systems a more general explanation in terms of vortex behaviour was developed [27]. A notable example of the zero field negative transverse resistance, similar to that shown in inset of figure 1 (a) is that of the negative anomalous Hall Effect (nAHE) which was observed in SrRuO$_3$ (due to the chiral nature of the superconducting state) and superconducting ferromagnets [28]. Additionally there is a large amount of evidence suggesting that a π-junction Josephson Effect can lead to negative currents [29]. The latter case can be treated as a non-trivial Josephson junction (JJ) array induced current offset. The tendency for the resistance to go negative as current is lowered is also observed in magnetoresistance (MR) data [Fig. 1(c), (d), inset].

We employ the standard analysis of the *log-log* plots of *V-I* measured at various temperature points to validate the existence of a BKT transition, details of this analysis is given in a subsequent section dedicated to current characteristics [see III.C]. We determine a BKT temperature of 1.3 K for sample B5 and 2.8 K for B1. Additionally we fit the $R(T)$ data to the following interpolation formula adapted from the Halperin Nelson scheme[30]:

$$\frac{R_N}{R(T)} = 1 + \left(\frac{2}{A} sinh\left(\frac{b}{\sqrt{t_c}}\right)\right)^2. \qquad (1)$$

A is a prefactor of order 1, $t_c = (T - T_{BKT})/T_{BKT}$, and *b* is a material parameter related to vortex core energy which is estimated to be 2.37 and 1.51 for sample B1 and B5, respectively represented as the ratio from the value predicted by the XY-model ($\mu/\mu_{XY}$). This interpolation formula is valid in the temperature range between mean field and BKT critical points, the fitting to data is shown in Fig.1 (b) [30]. One notable feature of the fitting is an underestimation of the resistance around the superconducting-normal transition, this has been attributed to pseudo-gap above transition point [30].

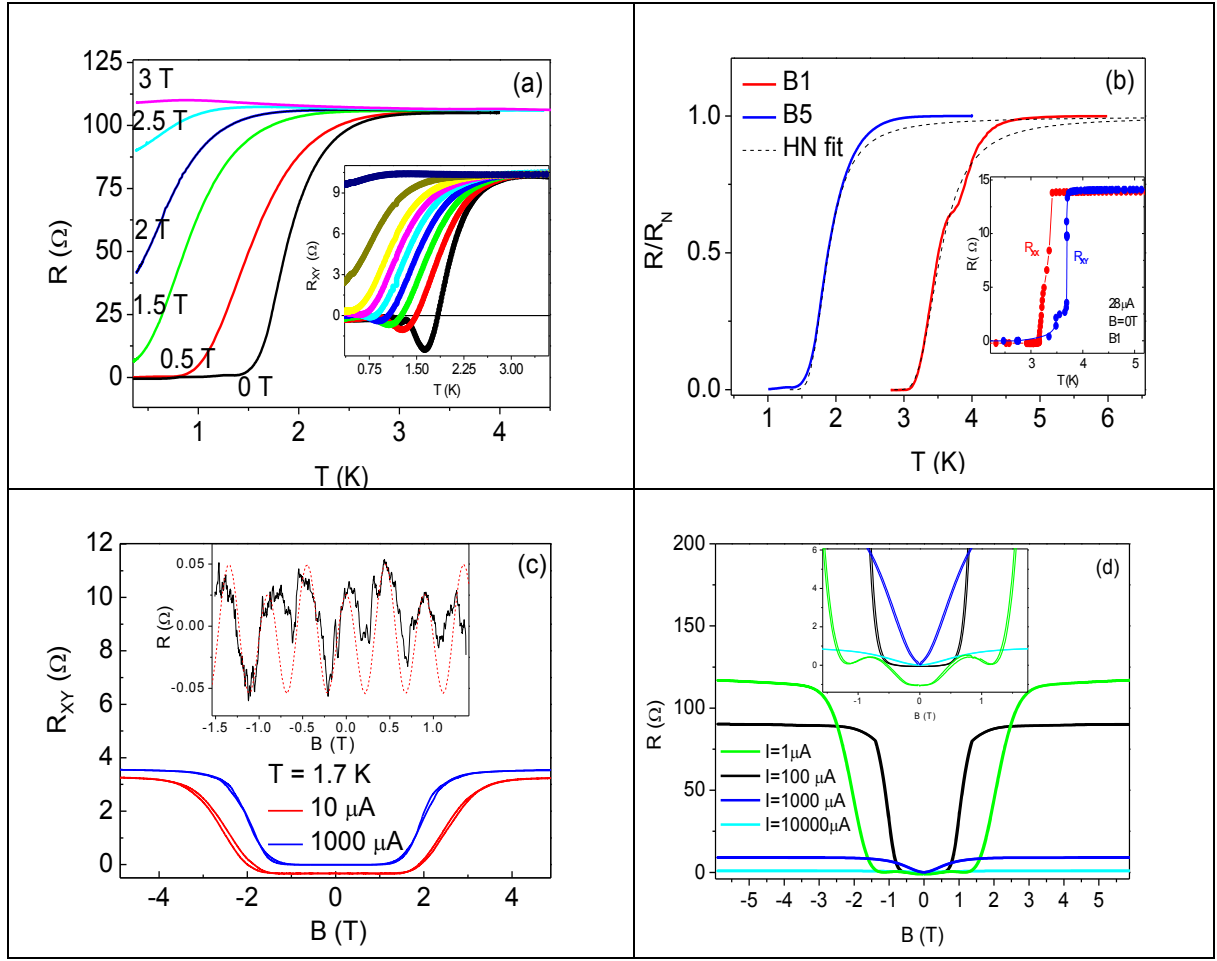

**FIG. 1(a)** Resistance as a function of temperature at different magnetic fields for sample B5. **Inset:** The transverse resistance shows a pronounced dip into an apparent negative regime, the dip is reduced upon application of magnetic field **(b)** Temperature dependent magnetoresistance normalized with respect to the normal state resistance, the solid line is a fit to the modified Halperin-Nelson formula. Under these measurement conditions sample B1 shows step feature in the fluctuation regime. **Inset:** $R_{XY}$-T and $R_{XX}$-T plots $B=$ 0 for sample B1 **(c)** $R_{XY}$ of sample B1 measured at different bias currents in the range of 10 µA to 1 mA within the magnetic field range of ± 4.5 T. Above 10 µA the superconducting region appears to be flat however at measurement current below 10 µA the resistance take on a finite negative value. **Inset:** Pronounced oscillatory features are observed in the low field region at current 1 µA, the dashed line is a fitting of Eqn. 2 after a background subtraction. **(d)** MR of sample B5 shows an increase in critical fields as the measurement current is increased. **Inset:** A zoomed in view shows a clear transition to a negative resistance regime at low measurement current.

Sample B5 also shows a deviation from the fit at the foot of the curve, this is most likely due to finite size effects [30] and is expected to be more pronounced than in sample B1 due to reduced grain size. These features can be seen as an indication that the microstructure plays an important role in the transport of these films. Interestingly the temperature resistance curves exhibit peak features at the BKT temperature (determined from the *I-V* characteristics), the peak height was found to increase in height when lowering the measurement currents.

**B. Magnetoresistance (MR) measurements**

MR measurements carried out at 0.3 K are shown in Fig. 1(c) and (d) where the observed features can be compared to transport in weakly disordered superconducting systems [20]. Both samples show an increase in MR from approximately ± 1.7 T at a current of 1 µA, levelling off around 3 T then changing little upon further increase of magnetic field. Upon reducing the current, sample B5 shows an increase in normal state resistance and the formation of a resonance like mode at $B = \pm$ 0.75 T (Fig. 1(d) and inset). The critical field is observed to increase with the reduction of the bias current 1(c) and 1(d). At low current additional features can be found in the region below the critical field that can be explained through an interference effect associated with the heterostructure of HBDDF, the following subsection explains the origin of such features. One of the most interesting features observed at low current regions (below 1 µA) are periodic oscillations in the superconducting region of the MR [Fig. 1(c), inset]. The oscillatory features appear to be different in sample B5 due to smaller grain size than sample B1 (explained below).

## MR at low current & low fields: Interference effects

Additional anomalous features were observed at low fields regions and also in the presence of current applied around or below the µA regime. These include, in addition to oscillatory behaviour, an insulator peak below $B_c$ on either side of zero magnetic field [Fig 1 (d) inset and Fig 2 (a)]. MR shows a hysteretic behaviour. Qualitatively, this means the resistance increases sharply (Fig. 2(a)) with increasing field up to a maximum (sharp peak) and then decreases again as field is further increased till it reaches the background curve. A temperature dependent crossover from peak maxima to a minima has been recorded (see Fig. 2(a), (b)). One feature is clear i.e. a transition from +ve to –ve MR at B→0 region. At 454 mK (and below) MR is –ve but it becomes +ve as the temperature rises. Without considering the hysteresis the magnetoconductance ($\Delta\sigma$) has been plotted in Fig. 2(b) which looks like a WL to WAL (weak anti-localization) transition. A similar behavior of WAL to WL transition has been explained by using the Hikami-Larkin-Nagaoka formula of magnetoconductivity [31]. The WAL effect can be explained by the Berry phase in a 2D Dirac fermion system expressed as $\phi_b = \pi(1-\Delta/2E_F)$ that clearly shows topological effect in these materials (depending on $\Delta$ and the Fermi level, $E_F$) [31]. A plausible mechanism of WL to WAL transition has been discussed in more detail elsewhere [32], here we relate it to the film microstructure (Fig. 2(c)).

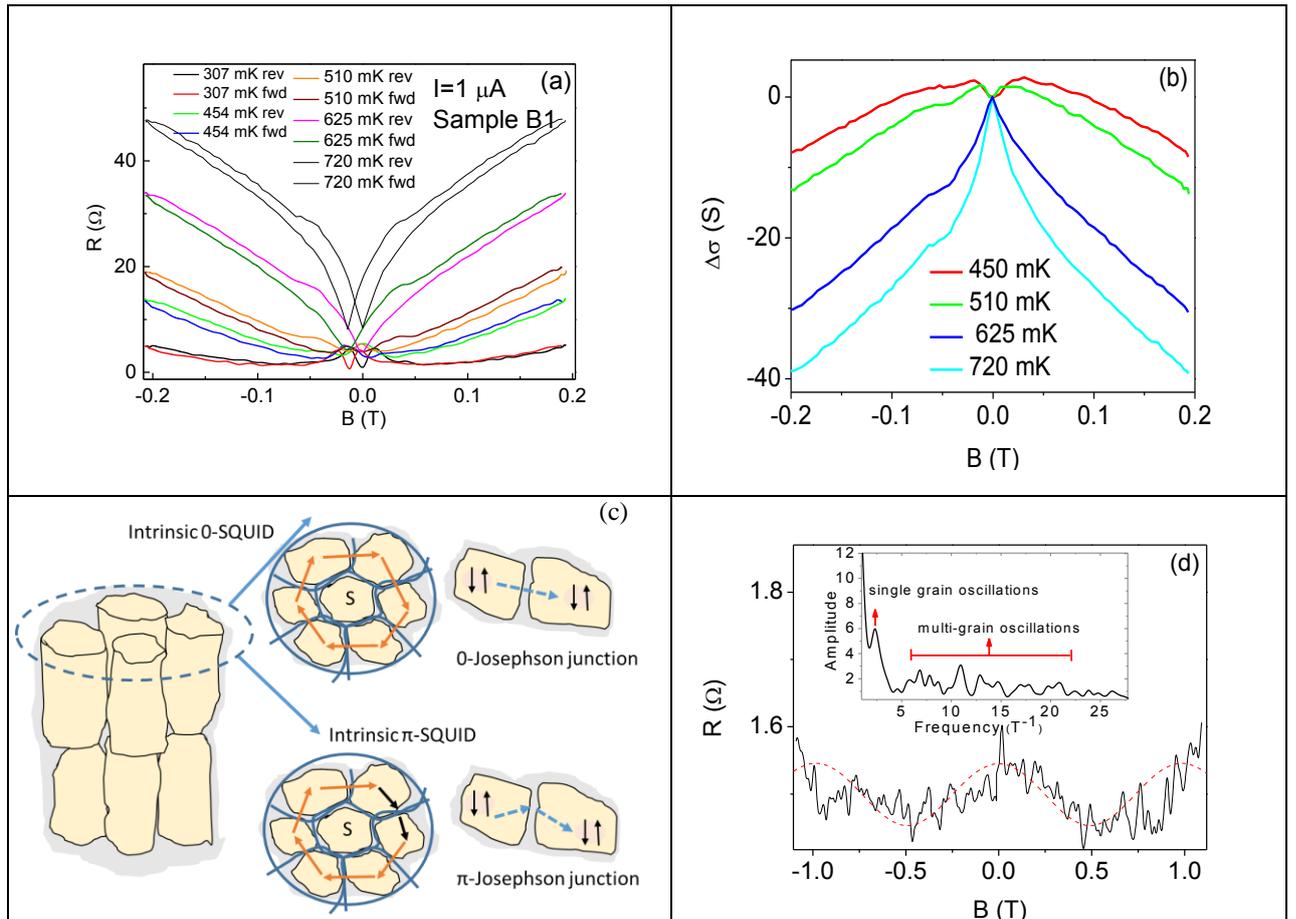

**FIG. 2(a)** Resistance as a function of *B* field recorded at various temperatures below the $T_c$ of sample B1. There is a clear transition from a positive to a negative magnetoresistance response as temperature is lowered, additionally there is a pronounced hysteresis effect. **(b)** The magnetoconductance data is plotted a conductance normalized to the zero field value to observe the change from positive to negative trend at the central part. **(c)** The proposed model of the microstructure describing the formation of a superconducting quantum interference pattern in nanodiamond films where 0-junction and π-junctions are formed due to different junctions. **(d)** Oscillations are recorded from sample B1 at low field (± 1 T) region. It clearly shows a MR peak arising at *B*= 0 point in addition to two other satellite peaks. This peak vanishes at higher currents e.g. 10 µA. Inset shows the fast FFT of the data with two different types of oscillations related to intergrain and intragrain structures.

By reducing the current to the nA range a drastic change in the low field MR is observed particularly a sharp peak centred at zero magnetic field that persists up to temperatures of 2.2 K. The central peak at *B*=0 appears from the insulating phase which is described by an S-I-S model of granular superconductors (Fig. 2(d)). These features can be attributed to the properties of intrinsic π-Josephson junctions as shown in Fig. 2(c) which consists of nanocrystals separated by thin grain boundaries. The hysteresis behaviour observed in the *R* vs. *B* plots at different temperatures below or around the $T_c$ (Fig. 2(a)) has not before been reported on in this system. It can be attributed to the strong inhomogeneity of the material bringing the hysteresis behaviour of two phase media which consists of strongly dissipative intergranular junctions with small critical currents and fields in addition to intra-granular effects. This concept is not novel, in fact Ji *et al*., analysed granular superconductor systems using two types of fluxons; (i) grain pinned fluxons and (ii) grain boundary fluxons [33]. Hence for granular superconductors two different critical current densities were expected. The critical current density within the grains is much greater than the intergrain region. At higher fields vortex clusters are formed and the fluxons are strongly pinned within the clusters. A similar kind of effect has been claimed by many other authors [34] [35] [36]. It should also be noted that similar transport phenomena has been observed in superconductors that exhibit negative transverse resistance [37] [38] [39]. However, we would like to extend this model for understanding of 0-π transitions in BNCD films.

If we accept the layered structure [in Ref. 24] of the superconductivity in the present HBDDF then a number of observed features can at least be qualitatively explained by a 0-π JJ hypothesis. The model of Spivac and Kivelson developed a few decades before [40] can be helpful in explaining our results in a qualitative manner. This model includes a resonant level and explained to be due to S-I-S structures. The key concept is based on explaining observations such as negative MR in terms of negative Josephson coupling. This model has been used for granular high $T_c$ materials in the vicinity of superconductor-insulator transition [40]. Although negative MR features have been reported by several researchers a detailed explanation has not yet been found [41] [42]. We believe that the negative resistance can also be related to the Josephson junctions where the negative superfluid density arises from the random distribution of coupling between grains in disordered media. The small oscillations in Fig. 2(d) we have recorded at a very low current near the low field region can be explained by the granular superconductor nanowire model shown in Fig. 2(c). There is a signature of phase slip of superconducting order parameter as suggested in 1D nanowires and d-wave superconductors [43-46].

In such a system one can imagine that the transport channel for Cooper pair bears a low dimensional network percolated character, in fact both negative MR and oscillations have before been reported for granular films which form nanowire paths in high $T_c$ [43-46]. In accordance with the current understanding, the quantum fluctuations of the order parameter in the superconducting region which is very common for 1D superconductors can be tracked through the corresponding fluctuating residual resistivity ($R$ vs. $T$) as we have observed in our measurements. These transport features can be explained by thermally activated phase slip (TAPS) or a quantum phase slip (QPS) which induce a transition from superconducting state to a normal state. Hence it adds a resistance to a superconductor even below $T_c$. Negative current and resistance can also arise from the anti-QPS phase [43-45]. It is well known that the Little-Parks theory cannot explain the MR and oscillations of large amplitude and this requires additional effect such as a BKT transition in specially decorated structures where the JJ is well defined. The proposed structure in Fig. 2(c) will produce a non-zero supercurrent, AB oscillations with a period of $h/4e$ and a negative MR. However, the period of oscillation of HBDDF films differs from the expected $h/4e$ which can be explained if one assumes the presence of natural π-SQUID and 0-SQUID structures connected in chains. Here we propose a microstructure model which consists of closed conduction loops through which flux is penetrating. This effectively leads to the appearance of the random distribution of the positive and negative supercurrents, this gives rise to a negative magnetoresistance contribution whereas the SQUID structures form an elementary unit allowing for the oscillatory behaviour of the magnetoresistance. In order to deeper investigate the properties of the oscillatory behaviour of the magnetoresistance we rely on the model developed in ref 46 that explains the oscillations arising from the coexistence of both π and 0 Josephson junctions effectively described by the following equation [46]:

$$V = \left(\frac{R1}{2}\right)\sqrt{I^2 - \left(2I_{c1}\cos\frac{\pi SH}{\phi_o}\right)^2} + \left(\frac{R2}{2}\right)\sqrt{I^2 - \left(2I_{c2}\sin\frac{\pi(2S)H}{\phi_o}\right)^2}, (2)$$

Where $I$, $I_{C1}$, and $I_{C2}$ {= 0.2 $I$} correspond to the measuring current and critical currents of the Josephson junction in 0 and π SQUIDs, respectively. $R_1$ and $R_2$ represent the single junction resistances and $S$ represents the effective area of 0-0 SQUID.

The oscillations are subjected to a Fourier transform to determine the magnetic period and consequently the dominant effective area. This is shown in the inset of figure 2 (d) (also in figure 1 (c)). It is observed that the dominant amplitude of oscillations is greatly dependent on the temperature, at temperatures around 2.6 K (i.e. within the superconducting regime) we find that the lower frequency oscillations (smaller orbits) are less pronounced. This suggests that the oscillatory behaviour of MR is due to closed paths formed by larger orbits comprised of multiple linked grains. As shown in the inset of figure 2(d) the oscillation amplitude are significant up to frequencies of 20 $T^{-1}$ corresponding to effective area with radius up to 117 nm, indicating SQUID loops of approximately three grains in radius, the best fit to data is obtained using the effective area obtained from the dominant peaks in the fast Fourier transform (FFT). As can be seen in figure 2(c) this leads a situation with a dominant 0-junction character with minimal π-junction behaviour. The qualitative features of the oscillations however change as the temperature is decreased. This is shown in the inset of figure 1 (c) where the best fit to the data ensures a dominant π -junction character. In this case the dominate FFT peaks are concentrated at the lower oscillations, these correspond to smaller orbits. The effective area of such orbits are found to correspond to the grain size of the sample. Indicating a cross over from multi-grain tunnelling to single grain transport regime which is also marked by an increase in the π-junction character. We can thus relate the anomalous MR features to the microstructure of the films where JJ arrays consisting of neighbouring grains having different values of order parameters (Fig. 2(c)). Such a non-equilibrium Josephson junctions can give an absolute negative resistance regime under the condition that the applied voltage should be less than the gap difference. We believe that the transport in sample B5 having significantly smaller grain size and the intergrain loop (*S*) than B1 suffers from π junctions which results in the observation of WAL effects, instead of WL (found in sample B1).

Recently, it has been shown that π-SQUID can be achieved on the basis of only geometrical and symmetry arguments [44,47]. Indeed, if the order parameter of the superconductor is anisotropic (as for *d*-type superconductors and probably as in our case due to observation of the Andreev Bound State) then, at some angles, ABS can contribute to the transport through the one junction of the SQUID leading to the π phase shift because the reflected particles suffer from the sign change of the pairing potential. At the same time the other junction of the SQUID can still be in a zero phase shift state due to different grain boundary (thicker) between its superconducting constituents. In a non-magnetic medium the π-junction is produced from strong Coulomb repulsion which create localized spins (charge Kondo-type) on a resonant impurity level. The possibility of charge Kondo resonance has been discussed elsewhere from a detailed analysis of re-entrance behaviour [32].

We started analysing our data through the random variations of *s*-wave order parameter to explain negative MR and magnetoresistance oscillations (MRO). We have established non-*s* wave symmetry of order parameter in this system. We thus evoke the existence of QPS, BKT transition and intrinsic π-junction arrays to explain our results [48]. In addition the non-*s* wave part can be explained by a triplet superconductor consisting of a ferromagnetic layer (F) and forming an S-F-S or F-S-F structure where the interface of F/S forms a π-junction and generates odd-frequency (triplet) pairing. This is confirmed from a spin-dependent phase shift [43]. The structure of BNCD film is similar to a *d*-wave superconductor where the sign change of the order parameter takes place on the granular surface. A random distribution of *d*-wave order parameter can produce a π junction [44-46]. This field requires a thorough investigation by angle dependent MR measurements and fitting to the MR data.

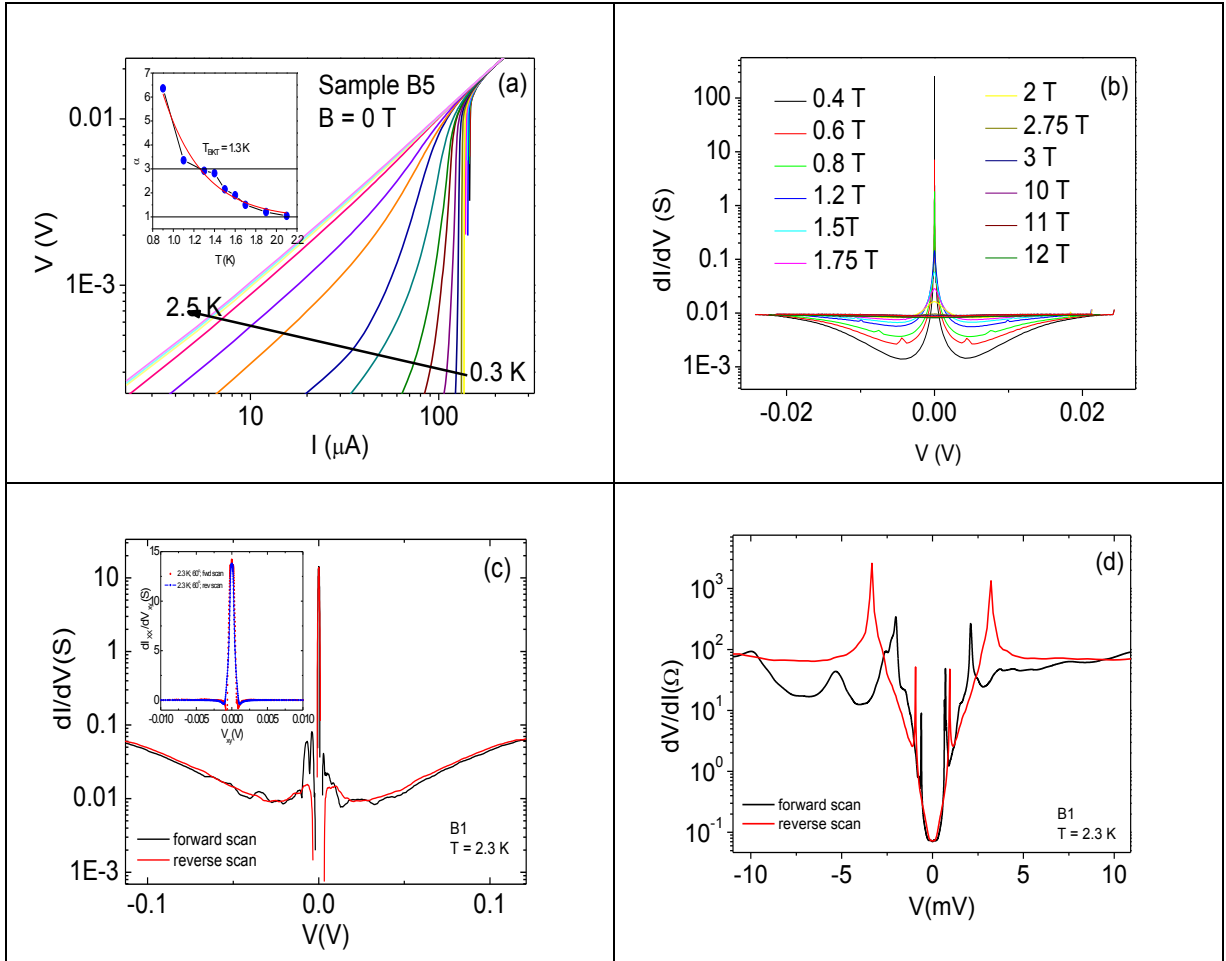

**Fig. 3 (a)** Voltage-current measurements at different temperatures ranging from 0.3 K to 2.5 K for sample B5. **Inset:** Power law fit to *I-V* data is used to verify the possibility for a BKT-like transition. **(b)** The corresponding differential conductance *vs.* voltage at 300 mK at various magnetic field 0 T to 12 T. A pronounced zero bias peak is observed that decreases in amplitude as the field is increased **(c)** The d*I*/d*V vs. V* plots corresponding to the same sample as in figure 3(c) shows the zero bias conductance anomaly, this time with additional peaks close to zero voltage. Inset shows the central part of the spectra a strong ZBCP feature plotted in the linear scale. **(d)** The forward and reverse measurement of the differential resistance as a function of applied voltage, the sharp peak features are related to the Andreev reflection process.

**C. Current-Voltage characteristics:**

**(i) BKT transition and Zero bias conductance peak**

In order to establish the BKT transition in the present system we make a detailed analysis of *I-V* characteristics (Fig. 3 (a) for sample B5) [49]. Here we investigate the *log-log* plot of the *I-V* curves in order to determine the power law dependence. This analysis is the so called KT-Nelson criterion and probes the superfluid stiffness which shows a jump at the BKT transition point [30]. The voltage shows a power law dependence on applied current, $V \sim I^{\alpha}$ where $\alpha = 1$ below the BKT transition and increase to above 3 past the critical point. From this analysis, shown in figures 3(a) we find a BKT temperature of approximately 1.3 K for sample B5. The estimated BKT temperature for sample B1 is 2.8 K. As sample B1 showed a higher mean field critical temperature a larger BKT temperature is expected. Our observations are consistent with other groups which provide a strong support for 2D vortices picture of the superconductivity in HBDDF. For example, Sacepe *et al.* established disordered Abrikosov lattice in a mixed state for HBDDF on the ground of the scanning tunnelling microscopy (STM) measurements [11]. Further analysis of the BKT transition involving the resistance temperature fitting using the Halperin Nelson equation has been presented in a previous section (III.A).

The differential conductance d$I$/d$V$ graphs are plotted as a function of applied voltage, these show a strong peak at zero current separated by two distinct minima for both samples B1 and B5 (Fig. 3(b), 3(c) and inset). With the increase of magnetic field the height of the ZBCP decreases while its width increases. Local zero bias conductance has before been observed through STM experiments and it was demonstrated that the order parameter had a special dependence at different values of the magnetic field at 0.37 K [3,11]. Indeed, it was established that superconductivity existed in all grains and moreover that the normal phase was also observed in the majority of grains. These observations have been linked to the strong inter- and intra-grain variations of the order parameter. Besides the ZBCP contribution, our conductance curves exhibit the same trend as was observed by Zhang *et al.* [3]. This explanation supports the idea that the global superconducting state with zero resistance emerges through the formation of percolation paths composed of grain-boundary Josephson junctions. Indeed, the presence of the peak can be attributed to the Andreev reflection from the surfaces of the boundaries which are naturally presented in the granular superconductors. The symmetry of the order parameter is crucial for observing such a pronounced ZBCP. Conventional *s*-wave superconductors do not exhibit the large ZBCP, therefore we can speculate that the symmetry of order parameter may not be *s*-wave type, however the exact nature of both symmetry and pairing (singlet-triplet) requires additional measurements. To verify the non-s wave character we plot the differential resistance corresponding to sample B1 (Figure 3 (d)). The sharp peak features within as well as near the superconducting gap are generally treated as evidence of the multiple Andreev reflection (MAR) where the position of such peak features are related to the interference of the Andreev currents generated due to the reflected particles [50]. These observations are also consists with the recent reports on non-*s* type symmetry of the order parameter for $MgB_2$ which was widely accepted as a reference material for modelling of the superconducting properties of the HBDDF [35] [36].

Having established BKT transition in these HBDDF we attempt to provide a qualitative description of different transition features observed in *R-T* and *R-B* measurements. Below the BKT transition temperature, vortices are bound to anti-vortices and preserve the superconducting coherence by allowing dissipation-less current to flow. With increasing magnetic field, these pairs become unstable and unbind resulting in free vortices which drift through the material leading to finite resistance and suppression of the BKT temperature. The Josephson coupling is strongest where diamond grains are closest together. HBDDF may therefore be interpreted as a disordered network of Josephson arrays. While the formation of vortices in boron-doped single crystal diamond has been clearly shown [11], the influence of the grain boundaries on vortex dynamics, which form a network through the entire material, is not yet known. HBDDF show some features in both MR measurements and *V-I* characteristics which are similar to features found in superconducting thin films of TiN [49]. These films forming superconducting islands and interact through Josephson coupling have also exhibited a BKT transition [22]. We understand that differential conductance peaks in Figure 3(b) and 3(c) still to be fitted or deconvoluted to explain the features. It does however indicate a mechanism that is in a competition with superconductivity and merits further investigation [32] [51].

## IV. CONCLUSIONS

HBDDF is considered to be a structurally complex system mixed with a disordered phase which shows some very interesting transport features including Andreev reflection, oscillatory magnetoresistance, a sharp transition at low fields and current induced metal-insulator transition around the superconducting transition point. These observations have not been reported elsewhere. In this article we explained the distinct local maximum in magneto-resistance measurements leading to a region of negative magneto-resistance through quantum interference effects which compete with superconductivity. Analysis of the oscillatory features of MR based on the interference effect finds non-s type order parameter due to formation of $S_1$-I-$S_2$ heterostructures. This analysis was found to be consistent with the microstructure of the diamond crystals where this granular media inherently forms 0 and π-Josephson junctions in a SQUID-like structure. The overall features are explained by the BKT transition in HBDDF since boron forms a 2D structure as observed recently. The superconductor-insulator transition has been attributed to a BKT transition where the corresponding magnetic field and the temperature have been determined from the *R-T* and *I-V* measurements. Since the theoretical work is limited in this (diamond) system [51] we presented a detailed experimental analysis which sheds light in understanding the superconducting transition in HBDDF and together with different effects [32] [46]. To the best of our knowledge a thorough investigation of the phase diagram of the superconducting state of the HBDDF is an open question. Therefore, the answer on the question about the particular symmetry of the order parameter of our samples requires further investigation. Nevertheless, on the ground of the presented set of data we can speculate about the exact symmetry of the assumed non-*s* pairing in the HBDDF.

## Acknowledgements

SB is very thankful to Prof. Miloš Nesládek for providing the diamond samples and P. Sheng (HKUST), A. Taraphder (IIT-KGP), J. Martinis (UCSB), and T. Okuma (TIT) for valuable comments on this work and also acknowledges CSIR-NLC, the URC Wits and National Research Foundation for the Nanotechnology Flagship Project for funding this project.
*Email: somnath.bhattacharyya@wits.ac.za

## References

[1] G. Zhang, T. Samuely, J. Kačmarčík, E. A Ekimov, J. Li, J. Vanacken, P. Szabó, J. Huang, P. J Pereira, D. Cerbu, and V. V. Moshchalkov, Phys. Rev. Applied **6**, 064011 (2016); G. Zhang, T. Samuely, Z. Xu, J. K Jochum, A. Volodin, S. Zhou, P. W. May, O. Onufriienko, J. Kačmarčík, J. A. Steele, J. Li, J. Vanacken, J. Vacík, P. Szabó, H. Yuan, M. BJ Roeffaers, D. Cerbu, P. Samuely, J. Hofkens, and V. V. Moshchalkov, ACS Nano **11**, 5358 (2017); N. Dubrovinskaia, R. Wirth, J. Wosnitza, T. Papageorgiou, H. F. Braun, N. Miyajima, and L. Dubrovinsky, Proc. Natl. Acad. Sci. **105**, 11619 (2008).

[2] E. Bustarret, Physics C **514**, 36 (2015); E. Bustaret, P. Achatz, B. Sacepe, C. Chapelier, C. Marcenat, L. Ortega, and T. Klein, Phil. Trans. R. Soc. A **366,** 267 (2008).

[3] G. Zhang, S. Turner, E. A. Ekimov, J. Vanacken, M. Timmermans, T. Samuely, V. Sidorov, S. M. Stishov, Y. Lu, B. Deloof, B. Goderis, G. Van Tendeloo, J. Van de Vondel, and V. V Moshchalkov, Adv. Mater. **26**, 2034 (2013).

[4] G. Zhang, M. Zeleznik, J. Vanacken, P. May, and V. Moshchalkov, Phys. Rev. Lett. **110**, 077001 (2013).

[5] H. Okazaki, *et al.*, Appl. Phys. Lett. **106**, 052601 (2015); Y. Takano, M. Nagao, I. Sakaguchi, M. Tachiki, T. Hatano, K. Kobayashi, H. Umezawa, and H. Kawarada, Appl. Phys. Lett. **85**, 2851 (2004).

[6] K. Ishizaka, R. Eguchi, S. Tsuda, a. Chainani, T. Yokoya, T. Kiss, T. Shimojima, T. Togashi, S. Watanabe, C.-T. Chen, Y. Takano, M. Nagao, I. Sakaguchi, T. Takenouchi, H. Kawarada, and S. Shin, Phys. Rev. Lett. **100**, 166402 (2008).

[7] W. Gajewski, P. Achatz, O. Williams, K. Haenen, E. Bustarret, M. Stutzmann, and J. Garrido, Phys. Rev. B **79**, 045206 (2009).

[8] Y. Tanaka, et al, J. Phys. Soc. Jpn. **71**, 2005 (2002); Y. Yanase and N. Yoroze, Sci. Technol. Adv. Mater. **9**, 044201 (2008); L. Boeri, J. Kortus, and O. K. Anderson Phys. Rev. Lett. **93**, 237002 (2004).

[9] F. Dahlem, P. Achatz, O. A. Williams, D. Araujo, H. Courtois, and E. Bustarret, Phys. Status Solidi A **207**, 2064 (2010); P. Achatz, W. Gajewski, E. Bustarret, C. Marcenat, R. Piquerel, C. Chapelier, T. Dubouchet, O. Williams, K. Haenen, J. Garrido, and M. Stutzmann, Phys. Rev. B **79**, 201203 (2009).

[10] T. Nishizaki, Y. Takano, M. Nagao, T. Takenouchi, H. Kawarada, and N. Kobayashi, Sci. Technol. Adv. Mater. **7**, S22 (2006).

[11] B. Sacépé, C. Chapelier, C. Marcenat, J. Kačmarčík, T. Klein, M. Bernard, and E. Bustarret, Phys. Rev. Lett. **96**, 097006 (2006).

[12] T. Shirakawa, S. Horiuchi, Y. Ohta, and H. Fukuyama, J. Phys. Soc. Japan **76**, 014711 (2007).

[13] G. Baskaran, J. Supercond. Nov. Magn. **21**, 45 (2007).

[14] J. J. Mareš, P. Hubík, M. Nesládek, D. Kindl, and J. Krištofik, Diam. Relat. Mater. **15**, 1863 (2006); J. J. Mares, P. Hubík, J. Krištofik, and M. Nesládek, Phys. Status Solidi **205**, 2163 (2008); J. J. Mares, P. Hubík, J. Kristofik, D. Kindl, and M. Nesládek, Chem. Vap. Depos. **14**, 161 (2008).

[15] S. Doniach (1984) '*Granular Superconductors and Josephson Junction Arrays*' in: Goldman A.M., Wolf S.A. (eds) Percolation, Localization, and Superconductivity. NATO Science Series (Series B: Physics), **109**. Springer, Boston, MA; S. Pace, R. De Luca, A. Saggese, Physica B **194**, 1551 (1994); Y. E. Lozoviktf and S. G. Akopov, J. Phys. C: Solid State Phys. **14** L31 (1981).

[16] Z.K. Tang, L. Zhang, N Wang, X.X. Zhang, G.H. Wen, G.D. Li, J.N. Wang, C.T. Chan, and P. Sheng, Science **292**, 2462 (2001).

[17] Z. Wang, W. Shi, H. Xie, T. Zhang, and N. Wang, Phys. Rev. B **81**, 174530 (2010).

[18] B.M. Kessler, C.O. Girit, A. Zettl, and V. Bouchiat, Phy. Rev. Lett. **104**, 047001 (2010).

[19] J. Vucicevic, M. O. Goerdig, and M. V. Milovanovic, Phys. Rev. B **86**, 214505 (2012).

[20] M. V. Milovanovic, and S. Predin, Phys. Rev. B **86**, 195113 (2012).

[21] M. Yu. Kagan, V. Mitskan, and M. M. Korovushkin, Eur. Phys. J. B **88**, 157 (2015); J. Supercond. Nov. Magn. **29**, 1043 (2016).

[22] F. Mancarella, J. Fransson, and A. Balatsky Supercond. Sci. Technol. **29**, 054004 (2016).

[23] R. Arenal, P. Bruno, D. J. Miller, M. Bleuel, J. Lal, and D. M. Gruen, Phys. Rev. B **75**, 195431 (2007); K.J. Sankaran, N. Kumar, S. Dash, H.C. Chen, A. K. Tyagi, N. H. Tai, and I. N. Lin, Surface and Coatings Technol. **232**, 75 (2013); Y. Tzeng, S. Yeh, W. C. Fang, and Y. Chu, Sci Rep. **4**:4531 (2014), G. Chimowa, D. Churochkin, and S. Bhattacharyya, Europhys. Lett. **99**,

27004 (2012); K. V. Shah, D. Churochkin, Z. Chiguvare and S. Bhattacharyya, Phys Rev B **82**, 184206 (2010).

[24] S. N. Polyakov, V. N. Denisov, B. N. Mavrin, A. N. Kirichenko, M. S. Kuznetsov, S. Y. Martyushov, S. A. Terentiev, and V. D. Blank, Nanoscale Res. Lett. **11**:11 (2016).

[25] R. P. Barber, Jr., S-Y. Hsu, J. M. Valles, Jr., R. C. Dynes, and R. E. Glover, Phys. Rev. B. **73**, 134516 (2006).

[26] S. J. Hagen et al., Phys. Rev. B47, 1064 (1993);R. Ferrell, Phys. Rev. Lett. **68**, 2524 (1992).

[27] S. J. Hagen, C. J. Lobb, R. L. Greene, M. G. Forrester, and J. H. Kang, Phys. Rev. B **41**, 11 630 (1990).

[28] N. Nagaosa, J. Sinova, S. Onoda, A. H. MacDonald, and N. P. Ong, Rev. Mod. Phys. **82**, 1539 (2010).

[29] J. Nagel, D. Speer, T.Gaber, A.Sterck, R.Eichhorn, P. Riemann, K.Ilin, M.Siegel, D. Koelle, and R. Kleiner. Phys. Rev. Lett. **100**, 217001 (2008).

[30] L. Benfatto, C. Castellani, and T. Giamarchi, Phys. Rev. B **80**, 214506 (2009); T. Schneider and S. Weyeneth, Phys. Rev. B **90**, 064501 (2014).

[31] H-Z Lu and S-Q Shen, Phys. Rev. B **92**, 035203 (2015); ibid Phys. Rev. B **84**, 125138 (2011); S. Hikami, A. Larkin, and Y. Nagaoka, Prog. Theor. Phys. **63**, 707 (1980).

[32] D. Mtsuko and S. Bhattacharyya, arXiv:1606.06672; C. Coleman, and S. Bhattacharyya, arXiv: 1706.02251.

[33] L. Ji M. S. Rzchowski, N. Anand, and M. Tinkham, Phys. Rev. B. **47**, 470 (1993).

[34] P. Vasek, I. Janecek, and V. Plechacek, Physica C **247**, 381 (1995); P. Vasek, Phys. Stat. Sol. (C) **3**, 3096 (2006).

[35] C. A. M. dos Santos, M. S. da Luz, B. Ferreira, and A. J. S. Machado, Physica C **391**, 345 (2003).

[36] A. Altinkok, K. Kilic, M. Olutas, A. Kilic, J. Supercond. Nov. Magn. **26**, 3085 (2016); B. Soodchomshom, I-M. Tang, and R. Hoonsawat, Physica C **468**, 47 (2008).

[37] P. A. Sobocinski, P. L. Grande, and P. Pureur, Physica C **517**, 48 (2015).

[38] T. L. Francavilla, E. J. Cukauskas, L. H. Allen, and P. R. Broussard, IEEE Transactions on App. Sup. **5**, 1717 (1995).

[39] H. Q. Nguyen, S. M. Hollen, M. D. Stewart, Jr., J. Shainline, A. Yin, J. M. Xu, and J. M. Valles, Jr., Phys. Rev. Lett. **103**, 157001 (2009); S. Ooi, T. Mochiku, and K. Hirata, Phys. Rev.Lett. **88**, 247002 (2002).

[40] B.Z. Spivac and S. A. Kivelson, Phys Rev B **43**, 3740 (1991); S. A. Kivelson and B.Z. Spivak, Phys Rev B **45**, 10490 (1992).

[41] Y. Chen, S. D. Snyder, and A. M. Goldman, Phys. Rev. Lett. **103**, 127002 (2009).

[42] A. V. Herzog, P. Xiong, and R. C. Dynes, Phys. Rev. B **58**, 14199 (1998).

[43] A. Bezryadin, J. Phys.: Cond. Matter **20**, 043202 (2008).

[44] A. Gumann, C. Iniotakis, and N. Schopohl, Appl. Phys. Lett. **91**, 192502 (2007), R. R. Schulz, B. Chesca, B. Goetz, C. W. Schneider, A. Schmehl, H. Bielefeldt, H. Hilgenkamp, J. Mannhart, and C. C. Tsuei, Appl. Phys. Lett. **76**, 912 (2000).

[45] S. Kawabata, S. Kashiwaya, Y. Tanaka, A. A. Golubov, and Y. Asano, J. Phys. Conf. Series **248**, 012039 (2010).

[46] D. Levi, A. Shaulov, A. Frydman, G. Koren, B. Ya. Shapiro, and Y. Yeshurun, EPL, **101**, 67005 (2013); D. Levi, A. Shaulov, G. Koren, and Y. Yeshurun, Physica C **495**, 39 (2013).

[47] D.A. Wollman, D.J. Van Harlingen, W.C. Lee, D.M. Ginsberg, and A.J. Leggett, Phys. Rev. Lett. **71**, 2134 (1993); J.H. Xu, J.H. Miller, Jr., and C.S. Ting, Phys. Rev. Lett. **51**, 11958 (1995); D. Litinski, M. S. Kesselring, J. Eisert, and F. von Oppen, Phys. Rev. X **7**, 031048 (2017).

[48] U. Patel, Z. L. Xiao, A. Gurevich, S. Avci, J. Hua, R. Divan, U. Welp, and W. K. Kwok, Phys. Rev. B **80**, 012504 (2009); P. A. Rosenthal, M. R. Beasley, K. Char, M.S. Colclough, and G. Zaharchuk, Appl. Phys. Lett. **59**, 3482 (1991); A. Pal, J.A. Ouassou, M. Eschrig, J. Linder, and M.G. Blamire, Sci. Rep. **7**, 40604 (2017).

[49] T. I. Baturina, S. V. Postolova, A. Y. Mironov, A. Glatz, M. R. Baklanov, and V. M. Vinokur, EPL **97**, 17012 (2012).

[50] G.E. Blonder, M. Tinkham, T.M. Klapwijk, Phys. Rev. B **25**, 4515 (1982); M. Tinkham, Introduction to Superconductivity (2d ed., McGrawHill, New York, (1996).

[51] R. McIntosh, N. Mohanto, A. Taraphder, and S. Bhattacharyya, arXiv:1509.09248v3.